\documentclass[twocolumn,showpacs,preprintnumbers,eqsecnum]{revtex4}
\usepackage{amsmath}
\usepackage{makeidx}
\usepackage{graphicx}
\usepackage{dcolumn}
\usepackage{bm}
\usepackage{hyperref}

\begin{document}

\preprint{APS/123-QED}
\title{The total cross-sections for the photoeffect for 2S-subshell bound
electrons and \\
pair production with the created electron in the 2S subshell for \\
photon energies above 1 MeV }
\author{A. Costescu$^{1}$,S. Spanulescu$^{1,2}$,C. Stoica$^{1}$}
\affiliation{$^{1}$Department of Physics, University of Bucharest, MG11,
Bucharest-Magurele 76900,Romania\\
$^{2}$Department of Physics, Hyperion University of Bucharest, Postal code
030629, Bucharest, Romania}

\begin{abstract}
Considering the contributions of the main term of the relativistic
Coulombian Green function given by Hostler [J. Math. Phys, \textbf{5}, 591
(1964)] to the second order of the S-matrix element and taking into account
only the large components of Dirac spinor of the 2S subshell, we have
obtained the imaginary part of the Rayleigh amplitudes in terms of
elementary functions. Thereby, simple and high accurate formulae for the
total cross-sections for photoeffect and pair production with the electron
created in the 2S subshell are obtained \textit{via} the optical theorem.
Comparing the predictions given by our formulae with the full relativistic
numerical calculations of Kissel et al [Phys. Rev. A 22, 1970 (1980)] and
Scofield [LLRL, Internal Report (1973)], a good agreement is found for low
and intermediate Z values if screening effects are taken into account, for
photon energies above the pair production threshold energy. We present our
numerical results for the total photoeffect and pair production
cross-sections for the 2S subshell of Zn and Ag atoms, for various photon
energies.
\end{abstract}

\maketitle

\section{\label{sec:level1}Introduction}

The S matrix element for Rayleigh scattering by 2S bound electrons of an
initial photon with momentum $\overrightarrow{k_{1}}=\omega \overrightarrow{%
\nu }_{1}$and polarization vector $\overrightarrow{s}_{1}$and a final photon
with momentum $\overrightarrow{k_{2}}=\omega \overrightarrow{\nu }_{2}$and
polarization vector $\overrightarrow{s}_{2}$ is:%
\vspace{-0.3 cm}%

\begin{equation}
\mathcal{M}=M\left( \omega ,\theta \right) (\overrightarrow{s}_{1}%
\overrightarrow{s}_{2})+N\left( \omega ,\theta \right) (\overrightarrow{s}%
_{1}\overrightarrow{\nu }_{2})(\overrightarrow{s}_{2}\overrightarrow{\nu }%
_{1})  \label{M}
\end{equation}

where $\theta $ is the photon scattering angle, and the invariant amplitudes 
$M$ and $N$ are:

\begin{eqnarray}
M(\omega ,\theta ) &=&\mathcal{O}-P\left( \Omega _{1},\theta \right)
-P\left( \Omega _{2},\theta \right)  \label{mn} \\
N(\omega ,\theta ) &=&Q\left( \Omega _{1},\theta \right) +Q\left( \Omega
_{2},\theta \right)  \notag
\end{eqnarray}

In the following we give the expression of the amplitude  $P(\Omega ,\theta ) $ which is the only term which has an imaginary in the case of forward scattering:

\begin{widetext}

\begin{eqnarray}
P(\Omega ,\theta ) &=&\frac{\omega \pm \omega _{pp}}{2m}\Bigg\{\frac{\varepsilon
^{5}\lambda ^{2}}{8m^{2}\omega ^{2}}\left[ \frac{\alpha Z\eta \Omega
-4\left( \eta ^{2}\mp \varepsilon m\omega \right) }{\left( 1+\frac{%
\overrightarrow{\Delta }^{2}}{4\eta ^{2}}\right) ^{2}}+\frac{2\left( \eta
^{2}\mp \varepsilon m\omega \right) +\frac{\overrightarrow{\Delta }^{2}}{%
\varepsilon }\frac{\Omega }{m}}{\left( 1+\frac{\overrightarrow{\Delta }^{2}}{%
4\eta ^{2}}\right) ^{3}}\right] \notag \\
&&+16\lambda ^{5}X^{2}\left[ \varepsilon ^{2}+\left( 1+\frac{\tau ^{2}}{4}%
\right) \frac{\alpha ^{2}Z^{2}}{4}\frac{X^{2}}{\omega ^{2}}-\frac{\alpha
^{2}Z^{2}\epsilon \Omega }{2m\omega ^{2}}\left( \eta ^{2}\mp \varepsilon
m\omega \right) \right] \frac{X}{d^{4}\left( \Omega \right) }\frac{%
F_{1}\left( 2-\tau ,2,2,3-\tau ;x_{1},x_{2}\right) }{2-\tau }  \notag \\
&&+32(1-\gamma )\lambda ^{5}X^{4}\left[\eta ^{2}+\left(1-\frac{1+\gamma}{2}\frac{ X^{2}}{\omega^{2}}-\frac{1-\gamma}{2}\frac{1-2\gamma}{2}\frac{\Omega^{2}}{\omega^{2}}\right)\overrightarrow{\Delta }
^{2}\right]\frac{X}{d^{6}\left( \Omega \right) }\frac{%
F_{1}\left( 3-\tau ,3,3,4-\tau ;x_{1},x_{2}\right) }{3-\tau }
\notag \\
&&+96\left( 1-\gamma \right) ^{2}\lambda ^{5}\frac{X^{6}}{\omega ^{2}}\Omega
^{2}\overrightarrow{\Delta }^{4}\frac{X}{d^{8}\left( \Omega \right) }\frac{%
F_{1}\left( 4-\tau ,4,4,5-\tau ;x_{1},x_{2}\right) }{4-\tau } \Bigg\}  
\label{pl}
\end{eqnarray}

\end{widetext}

with the parameters:%

\begin{eqnarray}
\Omega _{1}=\gamma m+\omega ,\Omega
_{2}=\gamma m-\omega =-\left\vert \Omega _{2}\right\vert ,
\end{eqnarray}
\vspace{0.2cm}
$\hspace*{1.1cm}\gamma=\left( 1-\alpha ^{2}Z^{2}\right) ^{\frac{1}{2}},\lambda =\alpha
Zm,\omega _{pp}=(1+\gamma )m , \\
\hspace*{1.3cm} \eta=\frac{\lambda }{2\epsilon }\;,\;\epsilon =\sqrt{\frac{1+\gamma }{2}},\tau\!=\frac{\alpha Z\Omega}{X},\overrightarrow{\Delta }=\overrightarrow{k_{1}}-\overrightarrow{k_{2}} $

and
\begin{equation}
X^{2}=-\omega ^{2}\mp 2\epsilon m\omega +\eta ^{2}\text{ with Re[}
X]>0 \label{par}
\end{equation}

\vspace*{-0.1cm}

The function $\mathcal{O}$ is the atomic form factor, while $F_{1}\left(
a;b_{1},b_{2};c;x_{1},x_{2}\right) $ is the Appell hypergeometric function
of four parameters and two complex variables given by the relationships: 
\vspace{-0.3cm} 
\begin{equation}
x_{1}x_{2}=p \left( \Omega \right)=\left[ \frac{d^{\ast }\left( \Omega \right) }{d\left(
\Omega \right) }\right] ^{2}=\xi ^{2}  \label{sum}
\end{equation}
\vspace{-0.6cm} 
\begin{equation}
x_{1}+x_{2}=s\left( \Omega \right)=2\xi -\frac{16X^{2}\omega ^{2}\sin ^{2}\frac{\theta }{2}}{%
d^{2}\left( \Omega \right) } \label{prod}
\end{equation}

with
\begin{eqnarray}
d\left( \Omega \right)=2\left( \eta ^{2}\mp \epsilon m\omega +\eta
X\right)  \label{num}\\
d^{\ast }\left( \Omega \right)=2\left( \eta ^{2}\mp \epsilon m\omega -\eta
X\right) \label{numst}
\end{eqnarray}

In the equations (\ref{par})-(\ref{numst}) the upper sign
corresponds to the case $\Omega =\Omega _{1}$, while the lower sign
corresponds to the case $\Omega =\Omega _{2}.$

We want to point out that the first term in the equation (\ref{pl}) which has no transcendental function and always real expression comes from reccurence relationships among Appell's functions, having no connection with the atomic factor.

Actually, using one more reccurence relationship, the equation (\ref{pl}) may be expressed in terms of only two independent Appell's functions.

According to the optical theorem, the imaginary part of the Rayleigh amplitude for forward scattering allows to get the total photoeffect and pair production cross-sections~:%
\vspace{-0.3 cm}
\begin{eqnarray}
\sigma _{ph} &=&\frac{4\pi }{\alpha }\frac{m}{\omega }r_{0}^{2}\left\vert 
\text{Im}[P\left( \Omega _{1},0\right) ]\right\vert  \label{sph} \\
\sigma _{pp} &=&\frac{4\pi }{\alpha }\frac{m}{\omega }r_{0}^{2}\left\vert 
\text{Im}[P\left( \Omega _{2},0\right) ]\right\vert
\end{eqnarray}%

\vspace{-0.3cm}

\section{The total cross-section of the photoelectric effect and pair
production in the case of 2s subshell}

Taking into account only the main term of the relativistic Coulombian Green
function given by Hostler\cite{ho} and considering only the large components
of the 2s subshell Dirac spinor, we obtain the imaginary part of the
amplitude for the forward elastic scattering of photons by 2s subshell
electrons \cite{cs} as follows:

\begin{widetext}%
\begin{align}
\left\vert \text{Im}[P\left( \Omega _{1},0\right) ]\right\vert & =2\frac{%
\omega +\omega _{pp}}{2m}\frac{\lambda ^{5}X_{1}^{2}\left( 1+\left\vert \tau
_{1}\right\vert ^{2}\right) \pi \left\vert \tau _{1}\right\vert \left\vert
X_{1}\right\vert }{6m^{4}\omega ^{4}}\frac{e^{-\left\vert \tau
_{1}\right\vert \chi _{1}}}{e^{\pi \left\vert \tau _{1}\right\vert }-e^{-\pi
\left\vert \tau _{1}\right\vert }}  \label{Imp1} \\
& \times \left\{ \frac{\eta ^{2}X_{1}^{2}}{m^{2}\omega ^{2}}\left( 1+\frac{%
\left\vert \tau _{1}\right\vert ^{2}}{4}\right) \left( 1-\frac{4\eta ^{2}}{%
5m^{2}}\right) -\left[ \epsilon ^{2}+\frac{\lambda ^{2}X_{1}^{2}}{%
2m^{2}\omega ^{2}}-\frac{\epsilon \alpha ^{2}Z^{2}\Omega _{1}}{2m\omega ^{2}}%
\left( \eta ^{2}-\varepsilon m\omega \right) \right] \right\}  \notag
\end{align}

\begin{align}
\left\vert \text{Im}[P\left( \Omega _{2},0\right) ]\right\vert & =2\frac{%
\omega -\omega _{pp}}{2m}\frac{\lambda ^{5}X_{2}^{2}\left( 1+\left\vert \tau
_{2}\right\vert ^{2}\right) \pi \left\vert \tau _{2}\right\vert \left\vert
X_{2}\right\vert }{6m^{4}\omega ^{4}}\frac{e^{-\left\vert \tau
_{2}\right\vert \chi _{2}}}{e^{\pi \left\vert \tau _{2}\right\vert }-e^{-\pi
\left\vert \tau _{2}\right\vert }}  \label{Imp2} \\
& \times \left\{ \frac{\eta ^{2}X_{2}^{2}}{m^{2}\omega ^{2}}\left( 1+\frac{%
\left\vert \tau _{2}\right\vert ^{2}}{4}\right) \left( 1-\frac{4\eta ^{2}}{%
5m^{2}}\right) -\left[ \epsilon ^{2}+\frac{\lambda ^{2}X_{2}^{2}}{%
2m^{2}\omega ^{2}}+\frac{\epsilon \alpha ^{2}Z^{2}\Omega _{2}}{2m\omega ^{2}}%
\left( \eta ^{2}+\varepsilon m\omega \right) \right] \right\}  \notag
\end{align}

where $\ \ \lambda =\alpha Zm,\eta =\frac{\lambda }{2\varepsilon },\omega
_{pp}=(1+\varepsilon )m,X_{j}=m^{2}-\Omega _{j}^{2},$with Re[$X_{j}]>0,\tau
_{j}=\frac{\alpha Z\Omega _{j}}{X_{j}},$ $j=1,2$


and%
\begin{equation}
\chi _{1}=\pi -2\arctan \frac{\eta \left\vert X_{1}\right\vert }{\varepsilon
m\omega -\eta ^{2}},\chi _{2}=\pi -2\arctan \frac{\eta \left\vert
X_{2}\right\vert }{\varepsilon m\omega +\eta ^{2}},\text{ for }\omega
>\omega _{pp}  \label{hietc}
\end{equation}

Observing that $\left\vert \tau _{j}\right\vert ^{2}\left\vert
X_{j}\right\vert ^{2}=\alpha ^{2}Z^{2}\Omega _{j}^{2}$ and $1-\frac{\eta ^{2}%
}{m^{2}}=\varepsilon ^{2}$ and using eqs.(\ref{sph})-(\ref{Imp2}) we get the
2S subshell photoeffect and pair production total cross-sections:

\begin{align}
\sigma _{ph}& =2r_{0}^{2}%
\frac{\pi ^{2}\alpha ^{5}Z^{6}}{3}\frac{\omega +\omega _{pp}}{\omega }\frac{%
\left( \left\vert X_{1}^{2}\right\vert +\alpha ^{2}Z^{2}\Omega
_{1}^{2}\right) }{\omega ^{2}}\frac{\Omega _{1}m}{\omega ^{2}}\frac{%
e^{-\left\vert \tau _{1}\right\vert \chi _{1}}}{e^{\pi \left\vert \tau
_{1}\right\vert }-e^{-\pi \left\vert \tau _{1}\right\vert }}
\label{sph final} \\
& \times \left[ \frac{\alpha ^{2}Z^{2}}{4\varepsilon ^{2}\omega ^{2}}\left(
\left\vert X_{1}^{2}\right\vert +\frac{\alpha ^{2}Z^{2}\Omega _{1}^{2}}{4}%
\right) \left( 1-\frac{\alpha ^{2}Z^{2}}{5\varepsilon ^{2}}\right)
+\varepsilon ^{2}-\frac{\lambda ^{2}}{2m^{2}}\frac{\left\vert
X_{1}^{2}\right\vert }{\omega ^{2}}-\frac{\alpha ^{2}Z^{2}\Omega
_{1}\varepsilon }{2m\omega ^{2}}\left( \eta ^{2}-\varepsilon m\omega \right) %
\right]  \notag
\end{align}

\begin{align}
\sigma _{pp}& =2r_{0}^{2}\frac{\pi ^{2}\alpha ^{5}Z^{6}}{3}\frac{\omega
-\omega _{pp}}{\omega }\frac{\left( \left\vert X_{2}^{2}\right\vert +\alpha
^{2}Z^{2}\Omega _{2}^{2}\right) }{\omega ^{2}}\frac{\left\vert \Omega
_{2}\right\vert m}{\omega ^{2}}\frac{e^{-\left\vert \tau _{2}\right\vert
\chi _{2}}}{e^{\pi \left\vert \tau _{2}\right\vert }-e^{-\pi \left\vert \tau
_{2}\right\vert }}  \label{spp final} \\
& \times \left[ \frac{\alpha ^{2}Z^{2}}{4\varepsilon ^{2}\omega ^{2}}\left(
\left\vert X_{2}^{2}\right\vert +\frac{\alpha ^{2}Z^{2}\Omega _{2}^{2}}{4}%
\right) \left( 1-\frac{\alpha ^{2}Z^{2}}{5\varepsilon ^{2}}\right)
+\varepsilon ^{2}-\frac{\lambda ^{2}}{2m^{2}}\frac{\left\vert
X_{2}^{2}\right\vert }{\omega ^{2}}-\frac{\alpha ^{2}Z^{2}\left\vert \Omega
_{2}\right\vert \varepsilon }{2m\omega ^{2}}\left( \eta ^{2}+\varepsilon
m\omega \right) \right]  \notag
\end{align}
\end{widetext}

Obviously, in the above formulae, the screening effects must be considered
because in a 2S state they are important. In a roughly screening model which
allows keeping the coulombian shape of the spinors a realistic approach is
to consider an effective charge $Z_{eff}=Z-2$ which has to be used in all
numerical calculations.

\section{Numerical results and conclusions}
\begin{figure*}
 \includegraphics[width=5in,keepaspectratio=true]{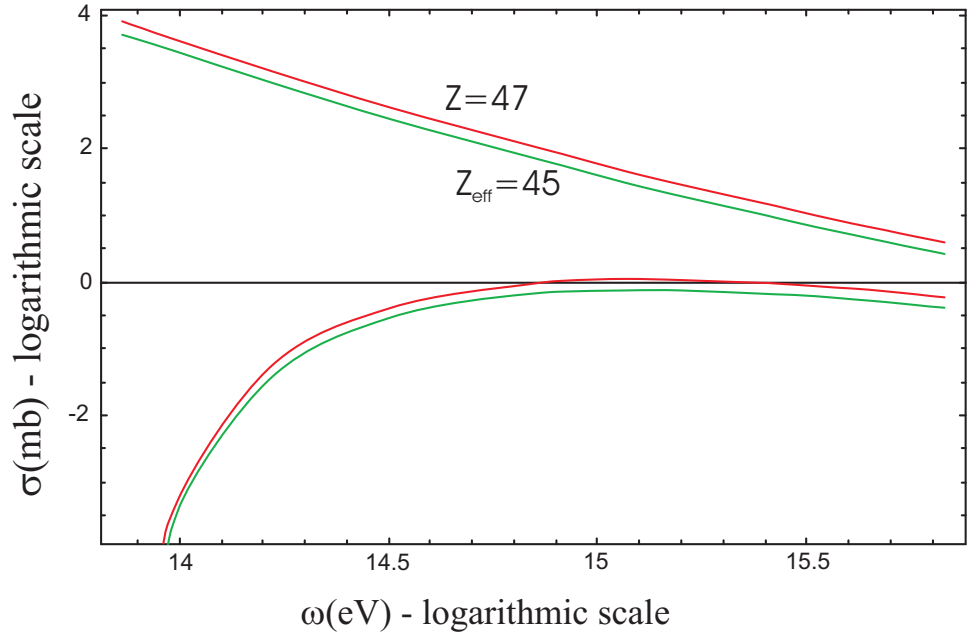}
   \caption{Photoeffect (upper) and pair production (lower) cross sections for 2S-subshell of Ag calculated for Z=47(red) respectively effective atomic number Z$_{eff}$=45(green).}
   \label{figura1}
\end{figure*}

\begin{figure*}
\includegraphics[width=5in,keepaspectratio=true]{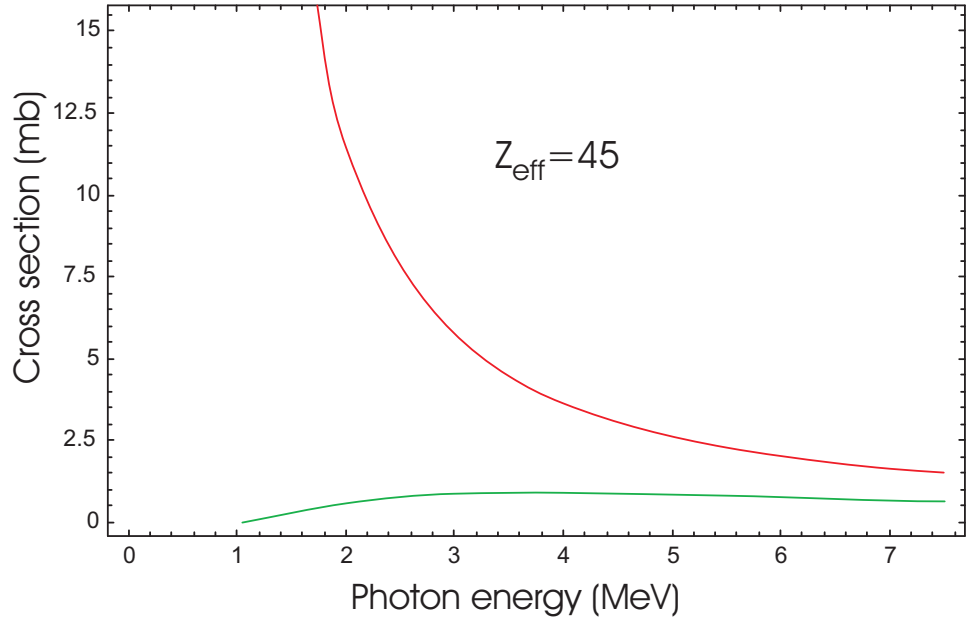}
   \caption{Photoeffect (red) and pair production (green) cross sections for Ag 2S subshell (Z$_{eff}$=45).}
   \label{figura2}
\end{figure*}

Using our analytical formulae for the cross sections for 2S subshell
electrons we get the numerical numerical  results in Table \ref{Zn}, Table %
\ref{Ag}, figure \ref{figura1} and figure \ref{figura2}. 

One may notice that changing the atomic number with it's effective value keeps the shape of the graphics, decreasing the whole spectrum values.
Also, it may be seen that the pair production cross section has a slight maximum, followed by a slow decreasing region. 

Comparing the predictions given by our formulae with the full relativistic
numerical calculations of Kissel \textit{et al} \cite{ki} and Scofield\cite%
{Sco},a good agreement within 10\% is found for low and intermediate Z
values if screening effects are taken into account, for photon energies
above the pair production threshold up to 5 MeV.
\vspace{0.3cm}
\begin{widetext}
\begin{table*}[!ht] %
\caption{Photoeffect and pair production cross sections for Zn 2S subshell(Z$_{eff}$=28).}%
\label{Zn}%
\begin{tabular}{r l  l  r @{.} l}
\hline\hline Energy (keV) \hspace*{10mm}
& Pair production &     Photoeffect &
\multicolumn{2}{l} {Cross sections}\\
\hspace{10mm} & cross section (mb) \hspace*{10mm}& cross section (mb)\hspace
*{10mm}&\multicolumn{2}{l}{ ratio}\\
\hline1022 \hspace*{10mm}  &  1.647x$10^{-10}$    &  5.95752      & 3&61x$10^{10}$\\
1100  \hspace*{10mm}  & 0.000695 &  5.07405   & 7298     & 68\\
1200  \hspace*{10mm}  & 0.00554 & 4.21312    & 759     &  343\\
1500  \hspace*{10mm}  & 0.036731 &   2.66716  &    72&61  \\
2000  \hspace*{10mm}  &  0.085664  &    1.54132 &  17&9925     \\
2754  \hspace*{10mm}  &  0.119357  &  0.881717   &    7&38719   \\
3000  \hspace*{10mm}  &  0.123683  & 0.765855    &  6&19209     \\
3500  \hspace*{10mm}  &   0.12716  &  0.599167  &  4&7119 \\
4000  \hspace*{10mm}  &  0.126442  &    0.488387 &     3&86253  \\
4807  \hspace*{10mm}  &   0.121266 &  0.372844   &    3&0746   \\
5000  \hspace*{10mm}  & 0.119682   & 0.352461    &     2&94497  \\
5500  \hspace*{10mm}  &  0.115336 &   0.308281  &   2&67289    \\
6000  \hspace*{10mm}  &  0.110871  &   0.273507  &   2&4669    \\
6500  \hspace*{10mm}  & 0.106464   &  0.245495   &  2&3059     \\
7000  \hspace*{10mm}  &  0.102209  &  0.222493   &   2&17684    \\
7500  \hspace*{10mm}  &  0.0981532  &   0.203295  &  2&0712     \\
\hline\end{tabular}%
\end{table*} 

\begin{table*}[!ht] %
\caption{Photoeffect and pair production cross sections for Ag 2S subshell (Z$_{eff}$=45).}%
\label{Ag}%
\begin{tabular}{r l  l  r @{.}   l}
\hline\hline Energy (keV) \hspace*{10mm}
& Pair production &     Photoeffect &
\multicolumn{2}{l} {Cross sections}\\
\hspace{10mm} & cross section (mb) \hspace*{10mm}& cross section (mb)\hspace
*{10mm}&\multicolumn{2}{l}{ ratio}\\
\hline1022 \hspace*{10mm}  &  2.10404x$10^{-8}$    &  43.9758      & 2&09x$10^{9}$\\
1100  \hspace*{10mm}  & 0.0035494 &     37.5046 &      10566&4\\
1200  \hspace*{10mm}  &  0.0329387 &     31.1846 &      946&746\\
1500  \hspace*{10mm}  & 0.242849 &     19.7968 &      81&519\\
2000  \hspace*{10mm}  & 0.586374 & 11.4659    &      19&554\\
2754  \hspace*{10mm}  & 0.826457 &     \texttt{~}6.56745 &      17&94651\\
3000  \hspace*{10mm}  & 0.857616 &      \texttt{~}5.70553 &      6&65278\\
3500  \hspace*{10mm}  & 0.883206 &     \texttt{~}4.46472 &      5&05513\\
4000  \hspace*{10mm}  & 0.878962 &     \texttt{~}3.63959 &      4&14078\\
4807  \hspace*{10mm}  & 0.8435 &    \texttt{~}2.77862   &    3&29416  \\
5000  \hspace*{10mm}  & 0.83255 &   \texttt{~}2.62671   &  3&15501    \\
5500  \hspace*{10mm}  & 0.80242 &    \texttt{~}2.2974    &    2&86309  \\
6000  \hspace*{10mm}  &  0.77140&    \texttt{~}2.03818  &   2&64218   \\
6500  \hspace*{10mm}  & 0.740752   & \texttt{~}1.82936     &   2&4696   \\
7000  \hspace*{10mm}  &  0.711145  &   \texttt{~}1.65788    &   2&33128   \\
7500  \hspace*{10mm}  &   0.682912 &    \texttt{~}1.51476  &  2&21809    \\
\hline\end{tabular}%
\end{table*}%

\end{widetext}
\vspace*{5cm}

%

\index{sco}

\end{document}